\documentstyle[12pt]{article}

\begin{document}
\topmargin 0pt
\oddsidemargin 0mm
\vspace{3mm}
\vspace{10mm}
\begin{center}
{\Large{\bf Multipole Amplitudes of Pion Photoproduction on Nucleons 
up to $2~GeV$ within Dispersion Relations and Unitary Isobar Model}}\\
\vspace{1cm}
{\large I.G.Aznauryan}\\
\vspace{1cm}
{\em Yerevan Physics Institute,
Alikhanian Brothers St.2, Yerevan, 375036 Armenia}\\
{(e-mail addresses:  aznaury@jlab.org, aznaur@jerewan1.yerphi.am)}\\
\vspace{5mm}
Abstract\\
\end{center}
\vspace{3mm}
Two approaches for analysis of pion photo- and electroproduction
on nucleons in the resonance energy region are checked at $Q^2=0$
using the results of GWU(VPI) partial-wave analysis
of photoproduction data. The approaches are based on 
dispersion relations and unitary isobar model.
Within dispersion relations good description of photoproduction
multipoles is obtained up to $W=1.8~GeV$. Within unitary isobar model,
modified with increasing energy by incorporation
of Regge poles, and with unified Breit-Wigner
parametrization of resonance contributions,
good description of photoproduction
multipoles is obtained up to $W=2~GeV$.   
\newline
\vspace{1cm}
PACS numbers: 13.60.Le, 14.20.Gk, 11.55.Fv, 11.80.Et
\vspace{10mm}
\renewcommand{\thefootnote}{\arabic {footnote}}
\setcounter{page}{1}
\section{Introduction}

It is known that dispersion relations (DR) and the unitary isobar model
(UIM) constructed on the basis of effective Lagrangian approach
in Ref. \cite{1} are widely used for analysis of pion
photo-and electroproduction data and for extraction from these data
information on $\gamma^* N \rightarrow N^*$ verteces.
In this paper our goal is to check these approaches at $Q^2=0$
using the results of GWU(VPI) partial-wave analysis
of pion photoproduction made from threshold to $W=2~GeV$
(Ref.  \cite{2,3} and SAID program). In the case of dispersion relations
the main goal is to find the energy region
of applicability of DR, because with increasing energy
the utilization of DR   
is connected with the following problems:
(a) at large angles the important
$P_{33}(1232)$ resonance contribution
requires in the integrands of DR
extrapolation to very large $x=cos\theta$
and becomes arbitrary,
(b) the unknown
contribution of resonances with large masses ($M~>~2~GeV$)
can become important,
(c) the contribution of the Regge
region ($W~>~2.5~GeV$) can also become important.
From the results of this paper it follows,
that the role of these effects is insignificant up
to $W=1.8~GeV$, and DR can be reliably
used in this energy region.

In the case of the UIM our goal is to find  adequate
description of the resonance
and background contributions in order  to extend
this model to $W=2~GeV$. It is known that the background
of the UIM  which consists 
of the minimal number of diagrams
(the nucleon exchanges in the $s$- and $u$-channels
and the $t$-channel $\pi$, $\rho$ and $\omega$ exchanges)
is  motivated  only at the 
threshold
(Refs.  \cite{4,5})
and in the first resonance region (Refs. \cite{1,6} and refferences
therein ).
As it will be argued in Section 3, the
extension of this background
above first resonance region can not be satisfactory.
Moreover,
continued to $2~GeV$ and higher the background of Ref. \cite{1}
strongly contradicts  experimental data.
We will modify
the UIM in such a way
that with increasing energy
the amplitudes of the model 
will transform into the amplitudes in the Regge pole 
regime
which starts practically at  $W=2.5~GeV$.
We will demonstrate that incorporation
of Regge poles into the background with increasing energy
results in
good description of GWU(VPI) photoproduction
multipole amplitudes in the resonance
energy region 
up to  $W=2~GeV$.
This description will be obtained using 
standard Breit-Wigner parametrization
for the resonance contributions suggested
in Ref.  \cite{7}.
Only  for the multipoles  $M_{1+}^{3/2},~E_{1+}^{3/2}$  corresponding
to the $P_{33}(1232)$ resonance a slight
modification will be
made in order to satisfy the Watson theorem  (\cite{8}).
Let us note that  
in order to reproduce
photoproduction multipole amplitudes
in the UIM  of Ref.  \cite{1} 
complicated parametrization
of resonance contributions has been  used.
Such complication 
is caused by the fact
that above first resonance region
background contributions
into some multipole amplitudes 
in  Ref.  \cite{1} become too large, and in order
to compensate  them resonance contributions
have been strongly deformed. 

In  Section 2 we will present our results obtained
within dispersion relations.
Firstly, in Sect. 2.1 we will present the results for multipole
amplitudes $M_{1+}^{3/2},~E_{1+}^{3/2}$ which correspond
to the $P_{33}(1232)$ resonance. It will be shown, that DR
for  $M_{1+}^{3/2},~E_{1+}^{3/2}$ can be transformed
into singular integral equations. These multipoles
will be found via solutions of these equations.
Further, in Sect. 2.2 imaginary parts of the contributions
corresponding to other resonances will be found
using the results of GWU(VPI) analysis and assuming Breit-Wigner
parametrization of resonance contributions.
We will find also nonresonance contributions
into imaginary parts of  $E_{0+}^{(0,1/2,3/2)}$, $M_{1-}^{(3/2)}$,
$M_{1+}^{(0,1/2)}$ and $E_{1+}^{(0,1/2)}$
which are not small at small energies due to
large $\pi N$ phases  $\delta_{0+}^{1/2,3/2}$, $\delta_{1-}^{3/2}$
and   $\delta_{1+}^{1/2}$. These contributions
will be found using DR
and the Watson theorem.
Finally, in Sect. 2.3 using DR,
real parts of the multipole amplitudes will be found.
In this Section the contribution of the multipole  $M_{1+}^{(3/2)}$
coresponding to the 
$P_{33}(1232)$ resonance into DR
for other multipoles
will be investigated, and
it will be shown that at $W>1.8~GeV$
there is significant arbitrariness in the real parts
of some multipole amplitudes connected with this contribution.
The dispersion integrals over high energy region
will be also estimated  in Sect. 2.3, and it will be shown
that the role of these integrals in the resonance energy 
region is insignificant.

In  Section 3 we will discuss and present
the modified UIM and will present the results
obtained within this model. 
Conclusions will be made
in Section 4.

\section{Dispersion relations}

We use fixed-t dispersion relations for invariant amplitudes.
Real parts of multipole amplitudes
we derive via expansion of obtained results over multipoles.  
The invariant amplitudes are chosen following the work  \cite{9}
in accordance with the following definition of the hadron 
electromagnetic current:
\begin{eqnarray}
I^\mu =\bar{u}(p_2)\gamma _5 \left\{ \frac{B_1}{2}\left[ \gamma
^\mu (\gamma \tilde{k})-(\gamma \tilde{k})\gamma ^\mu \right]+2P ^\mu
B_2+2\tilde{q}^\mu B_3+2\tilde{k}^\mu B_4\right.\nonumber \\
\left.-\gamma ^\mu B_5+(\gamma \tilde{k})P^\mu B_6+
(\gamma \tilde{k})\tilde{k}^\mu B_7+
(\gamma \tilde{k})\tilde{q}^\mu B_8\right\} u(p_1),
\label{1}\end{eqnarray}
where $\tilde{k},~\tilde{q},~p_1,~p_2$ are 4-momenta
of virtual photon, pion, initial and final nucleons, respectively, 
$P=\frac{1}{2}(p_1+p_2),~B_1,B_2,...B_8$
are invariant amplitudes which are functions
of the invariant variables 
$s=(\tilde{k}+p_1)^2,~t=(\tilde{k}-\tilde{q})^2,~Q^2\equiv 
-\tilde{k}^2$.

The conservation of the hadron electromagnetic current
leads to the relations:
\begin{eqnarray}
&&4Q^2B_4=(s-u)B_2 -2(t+Q^2-m_{\pi} ^2)B_3, \nonumber\\
&&2Q^2B_7=-2B'_5 -(t+Q^2-m_{\pi} ^2)B_8,
\label{2}\end{eqnarray}
where
$B'_5\equiv B_5-\frac{1}{4}(s-u)B_6$.
So, only the six of the eight invariant amplitudes
are independent. 
Let us choose as
independent amplitudes following ones:
$B_1,B_2,B_3,B'_5,B_6,B_8$.
The relations between these amplitudes  
and the multipoles
are given in Appendix A. 
For all amplitudes $B_1^{(\pm,0)},B_2^{(\pm,0)},B_3^{(+,0)},
{B'}_5^{(\pm,0)},B_6^{(\pm,0)},B_8^{(\pm,0)}$, except 
$B_3^{(-)}$,
unsubtracted dispersion relations at fixed $t$ can be written:
\begin{eqnarray}
Re~ B_i^{(\pm,0)}(s,t,Q^2)=&&R_i^{(v,s)}(Q^2)
\left(\frac{1}{s-m_N^2}+ 
\frac{\eta_i \eta^{(+,-,0)}}{u-m_N^2}\right)\nonumber\\
&&+\frac{P}{\pi }\int \limits_{s_{thr}}^{\infty}
Im~ B_i^{(\pm,0)}(s',t,Q^2)
\left(\frac{1}{s'-s}+ \frac{\eta_i\eta^{(+,-,0)} }{s'-u}\right) ds',
\label{3}\end{eqnarray}
where
$R_i^{(v,s)}(Q^2)$ are residues in 
the nucleon poles (they are given in Appendix B),
$\eta_1=\eta_2=\eta_6=1,~\eta_3=\eta'_5=\eta_8=-1$,
$\eta^{(+)}=\eta^{(0)}=1$, $\eta^{(-)}=-1$,
$s_{thr}=(m_N+m_{\pi})^2$.
For the amplitude $B_3^{(-)}$
we take the subtraction point at an infinity.
In this case using the current conservation
conditions (\ref{2}) we have
\begin{eqnarray}
Re~ B_3^{(-)}(s,t,Q^2)=&&R_3^{(v)}(Q^2)
\left(\frac{1}{s-m_N^2}+\frac{1}{u-m_N^2}\right)-
\frac{eg}{4\pi}\frac{F_\pi (Q^2) }{t-m_{\pi}^2 }\nonumber\\
&&+\frac{P}{\pi } \int \limits_{s_{thr}}^{\infty}
Im~ B_3^{(-)}(s',t,Q^2)
\left(\frac{1}{s'-s}+\frac{1 }{s'-u}-\frac{4 }{s'-u'}\right)ds',
\label{4}\end{eqnarray}
where the Born term, in addition to the nucleon
poles, includes also  the pion exchange in the 
$t$-channel.

Amplitudes $B^{(+)},~B^{(-)}$ correspond 
to isovector photon and are related
to the amplitudes with definite value of total isospin
in the $s$-channel by:
\begin{equation}
B^{(+)}=\frac{1}{3}(B^{(1/2)}+2B^{(3/2)}),
~~B^{(-)}=\frac{1}{3}(B^{(1/2)}-B^{(3/2)}),
\label{5}\end{equation}
amplitude $B^{(0)}$ corresponds to isoscalar photon.

Amplitudes corresponding to definite reactions are:
\begin{eqnarray}
&&B(\gamma~+~p\rightarrow p~+~\pi^0)=
B^{(+)}~+~B^{(0)},\nonumber \\
&&B(\gamma~+~n\rightarrow n~+~\pi^0)=B^{(+)}~-~B^{(0)},\\
&&B(\gamma~+~p\rightarrow n~+~\pi^+)=
2^{1/2}(B^{(0)}~+~B^{(-)}),\nonumber \\
&&B(\gamma~+~n\rightarrow p~+~\pi^-)=
2^{1/2}(B^{(0)}~-~B^{(-)}).\nonumber
\label{6}\end{eqnarray} 
We will use also the following notations:
\begin{equation}
{}_pB^{\frac{1}{2}}=B^{(0)}+\frac{1}{3}B^{(1/2)},
~~{}_nB^{\frac{1}{2}}=B^{(0)}-\frac{1}{3}B^{(1/2)}.
\label{7}\end{equation}
where the subscript $p~(n)$ denotes a proton (neutron) target.

\subsection{$P_{33}(1232)$ resonance}

Let us write dispersion relations for the multipole amplitudes  
$M_{1+}^{3/2}/kq,~E_{1+}^{3/2}/kq$ in the form
\begin{equation}
M(W)= M^B(W)+
\frac{1}{\pi}\int\limits_{W_{thr}}^{W_{max}}
\frac{ImM(W')}{W'-W-i\varepsilon}dW' 
+\frac{1}{\pi}\int\limits_{W_{thr}}^{W_{max}} K(W,W')ImM(W')dW',
\label{8}\end{equation}
 where $M(W)$ denotes any of multipoles 
$M_{1+}^{3/2}/kq,~E_{1+}^{3/2}/kq$,
$k$ and $q$ are the photon and pion 3-momenta in the c.m.s.,
 $M^B(W)$ is the contribution of the Born term into these
multipoles, $K(W,W')$ is a nonsingular kernel arising from the
u-channel contribution into the dispersion integral and the
nonsingular part of the s-channel contribution,  $W_{max}=1.8~GeV$.

 In the relations (\ref{8}) the following 
assumptions are made:
\newline
(a) For each  amplitude 
 $M_{1+}^{3/2}/kq,~E_{1+}^{3/2}/kq$ we neglect the contributions
of other multipoles
into dispersion integrals. By our
estimations they are small and do not affect our final results.   
These contributions are shown by thin dashed curves
in Figs. 1a,b. It is seen that up to $1.8~GeV$
they are practically equal to 0 for $M_{1+}^{3/2}$
and are very small for $E_{1+}^{3/2}$.
\newline
(b) We neglect the integrals over  ($W_{max}=1.8~GeV,~\infty$).
In Figs. 1a,b we have presented the results of GWU(VPI)
partial-wave analysis for
$M_{1+}^{3/2}$ and $E_{1+}^{3/2}$ up to   $W=2~GeV$.  
It is seen that  at  $1.8~GeV<W<2~GeV$, 
$ImM_{1+}^{3/2}$ and $ImE_{1+}^{3/2}$
are practically equal to 0.
The integrals over  ($2~GeV,~\infty$) were estimated using
the results of the
Regge-pole analysis of high energy data made in Ref. ~\cite{10}.
These integrals turned out to be negligibly small.

Let us make one more assumption:
\newline
(c) We will use the Watson theorem 
from threshold to  $W=1.8~GeV$, assuming that 
$\delta_{1+}^{3/2}(W)~\rightarrow~\pi$, i.e.
$ImM(W)\rightarrow 0$,
when $W~\rightarrow~W_{max}$.
From the results of GWU(VPI)
partial-wave analysis presented in
Figs. 1a,b it is seen that this is
reasonable assumption, taking into account
that the
$\pi N$ amplitude $h_{1+}^{3/2}(W)$
 is elastic up to $W=1.45~GeV$,
and  at  $W=1.45~GeV$,
 $\delta_{1+}^{3/2}=157^\circ$ \cite{11}.
So, in all integration region we will take
$ ImM(W)=h^*(W)M(W)$. 

With  these assumptions the dispersion relations for
 $M_{1+}^{3/2},E_{1+}^{3/2}$
 turn into singular integral equations (\ref{8}), 
which at $K(W,W')=0$ have
solutions in an analitical form (\cite{12,13}):
\begin{equation}
M_{K=0}(W)=M_{part,K=0}^{B}(W)+ 
c_MM_{K=0}^{hom}(W), 
\label{9}\end{equation} 
where
$M_{part,K=0}^{B}(W)$ is the 
particular solutions of Eq. (\ref{8}) generated by $M^B$:
\begin{equation}
M_{part,K=0}^{B}(W)=M^{B}(W)+
\frac{1}{\pi}\frac{1}{D(W)}
\int\limits_{W_{thr}}^{W_{max}}
\frac{D(W')h(W')M^{B}(W')}{W'-W-i\varepsilon}dW',
\label{10}\end{equation}
and
\begin{equation}
M_{K=0}^{hom}(W)=\frac{1}{D(W)}=
\frac{W_{max}}{|W_{max}-W|}\exp\left[\frac{W}{\pi}
\int\limits_{W_{thr}}^{W_{max}}\frac{\delta
(W')}{W'(W'-W-i\varepsilon)}dW'\right] 
\label{11}\end{equation} 
is
the solution of the homogeneous equation (\ref{8}) with $M^B=0$. It
enters the solution (\ref{9}) with an arbitrary weight, i.e.
multiplied by an arbitrary constant $c_M$.  

At $K(W,W')\neq~0$ one can transform the singular integral
equation~(\ref{8}) into nonsingular integral equation (\cite{14}).
The solution of this equation turned out
to be very close to (\ref{9}). 

From Eqs. (\ref{9}-\ref{11}) it is seen that with given  $M^B(W)$,
the particular and homogeneous 
solutions of the integral equation (\ref{8})
are determined only by the phase $\delta_{1+}^{3/2}(W)$.
In our calculations, at  $W<1.35~GeV$, 
this phase was taken in the form:
\begin{equation}
sin^2\delta_{1+}^{3/2}=
\frac{(4.27q^3)^2}{(4.27q^3)^2+(q_r^2-q^2)^2
[1+40q^2(q^2-q_r^2)+21.4q^2]^2},   
\label{12}\end{equation}
which is a slightly modified version of the corresponding
formula from Ref. \cite{15}, $q_r$ is the
pion c.m.s. momentum at $W=M$. At $1.35~GeV<W<1.8~GeV$,  the phase  
was taken in the form:
\begin{equation}
\delta=\pi-[\pi-\delta(W=1.35~GeV)]
\left(\frac{q-q_2}{q_1-q_2}\right)^2,   
\label{13}\end{equation}
where $q_1$ and $q_2$ 
are the pion c.m.s. momenta, respectively,
at $W=1.35~GeV$ and $W=1.8~GeV$.
Eqs.  (\ref{12},\ref{13}) reproduce with good accuracy
the results of the GWU(VPI) analysis 
for $\delta_{1+}^{3/2}(W)$ 
from threshold to  $W=1.5~GeV$.

Our final results for 
 $M_{1+}^{3/2},~E_{1+}^{3/2}$, obtained by the formulas
(\ref{9}-\ref{13}) 
via adjusting the only unknown parameters $c_M$ and $c_E$,
are presented in Figs. 1a,b.
It is seen that there is good agreement
with  the GWU(VPI) results.
In Fig. 1a we have presented also separately
contributions of particular and homogeneous solutions
into imaginary part of  $M_{1+}^{3/2}$.

\subsection{Imaginary parts of multipole amplitudes up to $2~GeV$}

In the resonance energy region imaginary parts
of multipole amplitudes are determined mainly by the resonance
contributions. These contributions for the resonances
 which are seen in the GWU(VPI) analysis at
$W<2~GeV$, except $P_{33}(1232)$, 
we have found using Breit-Wigner parametrization
given in Appendix C. Good description of the imaginary parts
of multipole amplitudes was obtained with parameters in the 
Breit-Wigner formula
presented in Table \ref{tab1}.

\begin{table}[t]
\begin{center}
\begin{tabular}{|c|c|c|c|c|}
\hline
Resonance&$M,~GeV$&$\Gamma,~GeV$&$X$\\
\hline
$P_{11}(1440)$&1.45 (1.46)&0.3&0.3\\
\hline
$S_{11}(1535)$&1.52&0.11 (0.12)&0.5\\
\hline
$D_{13}(1520)$&1.51&0.12 (0.1)&0.1 (0.3)\\
\hline
$S_{11}(1650)$&1.65&0.08 (0.11)&0.5\\
\hline
$F_{15}(1680)$&1.68&0.13&0.2 (0.5)\\
\hline
$P_{13}(1720)$&1.8&0.38 (0.35)&0.5 (0.4)\\
\hline
$S_{31}(1620)$&1.61 (1.62)&0.14&0.5\\
\hline
$D_{33}(1700)$&1.65&0.25&0.22 (0.2)\\
\hline
$F_{37}(1950)$&1.92 (1.93)&0.3&0.5\\
\hline
\end{tabular}
\caption{\label{tab1}Parameters in the Breit-Wigner formula
from Appendix C, found via description of imaginary parts
of multipole amplitudes. 
In the parentheses the parameters found within
UIM are presented in the cases when they differ from the parameters
found in this Section.}
\end{center}
\end{table}

At small energies imaginary parts of the amplitudes
$E_{0+}^{(0,1/2,3/2)}$, $M_{1-}^{(3/2)}$,
$M_{1+}^{(0,1/2)}$, $E_{1+}^{(0,1/2)}$
contain also noticeable nonresonance contributions
due to
large $\pi N$ phases  $\delta_{0+}^{1/2,3/2}$, $\delta_{1-}^{3/2}$
$\delta_{1+}^{1/2}$. 
Up to $W=1.3~GeV$ these contributions were found
using DR and the Watson theorem.
At higher energies they were 
reduced to 0.

Our final results for the imaginary parts of the multipole
amplitudes obtained in the way described in this 
Section are presented
by dashed curves in Fig. 2.

\subsection{Real parts of multipole amplitudes}

Real parts of multipole amplitudes were found using
dispersion relations  (\ref{3},\ref{4}).
Let us present dispersion integrals in these relations
in the form:
\begin{equation}
\int\limits_{W_{thr}}^{\infty}=
\int\limits_{W_{thr}}^{2~GeV}+
\int\limits_{2~GeV}^{2.5~GeV}+
\int\limits_{2.5~GeV}^{\infty}.
\label{14}\end{equation}

Imaginary parts of the amplitudes in the first
integral over resonance energy region from threshold to $2~GeV$
are known from the results of Sections 2.1 and 2.2.
The only uncertainty which  can arise in this integral
with increasing energy at large angles
is connected with the contribution of the $P_{33}(1232)$
resonance. 
This  uncertainty is the result of the  extrapolation of 
the  $P_{33}(1232)$ contribution in the integrands
at large $\mid t \mid$ to very large
$x'=cos(\theta')$. Such extrapolation with increasing $x'$
becomes arbitrary. In addition, the contribution of the multipole   
$M_{1+}^{3/2}$, 
which is large by itself, sharply grows with increasing energy.
This is demonstrated in Figs. 1b and 3 for the amplitudes
$E_{1+}^{3/2}$,  $E_{0+}^{1/2,3/2}$, $M_{1-}^{1/2,3/2}$.
Let us note, that at $W>1.7-1.8~GeV$
these amplitudes
are much smaller than the  $P_{33}(1232)$ resonance contribution
(see Figs. 2a-f), and therefore should be obtained
as differences of two large contributions,
one of which, the  $P_{33}(1232)$  contribution,
contains arbitrariness.
By this reason, with increasing energy we have 
significant arbitrariness
in the real parts of some multipole amplitudes
connected with  the  $P_{33}(1232)$ resonance contribution.
 
Imaginary parts of the amplitudes 
in the integrals
over high energy region
from $2.5~GeV$ to $\infty$ can be found using the results
of Regge-pole analysis made in Ref. \cite{10}.
The way of construction of the amplitudes in Ref.
\cite{10} is described in Appendix D. 
This analysis is made 
in gauge-invariant form and
in terms of invariant amplitudes. By this reason its
results can be easily used for the calculation 
of high energy integrals
in the dispersion relations (\ref{3},\ref{4}).
The role of these integrals turned out to be negligible in
the real parts of multipoles in the I and II resonance regions
and very small in the III and IV resonance regions.

Imaginary parts of the amplitudes in the integrals 
over intermediate energy region
were calculated via interpolation of the
imaginary parts of the amplitudes between the regions:
$W<2~GeV$ and $W>2.5~GeV$. 

Our final results for real parts of the multipole
amplitudes are presented  in Fig. 2
by dashed-dotted curves.
Let us note that in  Fig. 2 all amplitudes, for which the
GWU(VPI) analysis gives definite results,
are presented.   
It is seen that the agreement with the GWU(VPI) results
is satisfactory  up to $W=1.8~GeV$.
At higher energies in some multipoles noticeable
deviations from the GWU(VPI) results arise. Most probably they
are connected with the above discussed  
$P_{33}(1232)$ resonance contribution.
These deviations strongly depend on the small changes
of the magnitude and the shape of the multipole $M_{1+}^{3/2}$
corresponding to the $P_{33}(1232)$ resonance.
In addition, when we approach $W=2~GeV$, the  uncertainty
connected with
the unknown contribution of the resonances with
$M>2~GeV$ can become significant.
By these reasons
multipole amplitudes obtained by DR at
 $W>1.8~GeV$ are not reliable, and we do not present them in Fig. 2.

\section{Unitary isobar model}

The unitary isobar model is based on the effective Lagrangian
approach which was introduced in Refs. \cite{4,5} to reproduce
low energy results of current algebra and PCAC.
Within the
approach of  Refs. \cite{4,5} the  pion photoproduction
amplitudes consist of nucleon exchanges in the $s$-
and $u$-channels and $t$-channel
$\pi$ exchange
with pseudovector $\pi NN$ constant 
and contact term presented in Appendix B.
These contributions describe well the pion photoproduction on the nucleons
at the threshold ( Ref. \cite{5}). 

Later the approach of  Refs. \cite{4,5} was extended to the 
$P_{33}(1232)$ resonance region in the number of works
(see, for example, references in Refs. \cite{1,6}), 
and to the first, second and third
resonance regions in the UIM \cite{1}.
Background of  the UIM is constructed
from the contributions of the nucleon exchanges in the $s$-
and $u$-channels and $t$-channel
$\pi$ exchange 
with  $\pi NN$ constant which, being pure
pseudovector at the threshold, with increasing energy transforms
into pseudoscalar constant via the formula (B.2).
In addition to these contributions, the background 
of the UIM \cite{1}
contains the  $t$-channel
$\rho$ and $\omega$-exchanges,
which are given in Appendix E.
The background, constructed in this way,
is unitarized for each multipole amplitude 
according to  Ref. \cite{16}
in the $K$-matrix
approximation:
\begin{equation}
Unitarized(M_{l\pm},~E_{l\pm},~S_{l\pm})_{background}=
(M_{l\pm},~E_{l\pm},~S_{l\pm})_{background}(1+ih^{\pi N}_{\pm}).
\label{15}\end{equation}
With increasing energy the contribution
of this background into some multipole amplitudes
becomes too large. This is demonstrated in the case
of $M_{1+}^{3/2}$ in Fig. 1d (dotted curve).
In order to compensate these large background contributions,
resonance contributions
in the UIM
of Ref. \cite{1} have been strongly deformed.
Continued to the energies $W>2~GeV$
the background of Ref. \cite{1}
strongly contradicts experimental data.

Extension of the effective Lagrangian approach
above first resonance region with the minimal
set of diagrams (the nucleon exchanges in the $s$-
and $u$-channels and $t$-channel
$\pi$, $\rho$ and $\omega$-exchanges)
can not be satisfactory by the following reasons:

(i) Restriction by the mesons with lowest masses
in the $t$-channel exchanges is justified
only at small energies, where $t$ is small
and, therefore, the propogators $1/(t-m_{mes}^2)$
are determined by the meson masses.
However, with increasing energy the range of $t$
is increasing, and additional $t$-channel
contributions corresponing to the mesons with
higher masses should be taken into account.

(ii) With increasing energy, starting with $W=1.3~GeV$,
the contributions of inelastic channels into $\pi N$
scattering become important (see, for example, Ref. \cite{17}).
This means that the diagrams corresponding to the production
of other particles with subsequent rescattering, i.e.
$\gamma N\rightarrow inel. \rightarrow \pi N$,
should be taken into account.

So, in order to extent consistently effective Lagrangian approach
above first resonance region,
it is necessary to take into account
a large ammount of new diagrams.

From the other hand, it is known that with
increasing energy reggezation of different
contributions occurs via multiple gluon
exchanges between $t$-channel quarks,
and all contributions are reduced
to the restricted number of reggezied $t$-channel
exchanges. This picture, which is known as Regge-pole model,
gives good description of exclusive reactions
above $W=2.5~GeV$ ($E_{\gamma}=3~GeV$) at $t<3~GeV^2$.
In the number of cases Regge-pole approach gives good description
at smaller energies too.

By this reason we have modified the unitary isobar model
in such a way, that it incorporates
the results of the  effective Lagrangian approach
in the first resonance region and the Regge-pole behaviour
of amplitudes at high energies. With this aim we have
made the following modifications:
\newline
(a) Background is constructed in the form:
\begin{equation}
Back=[N+\pi+\rho+\omega]_{UIM}\frac{1}{1+(s-s_0)^2}+
[\pi+\rho+\omega+b_1+a_2]_{Regge}
\frac{(s-s_0)^2}{1+(s-s_0)^2},
\label{16}\end{equation}
where the Regge-pole amplitudes 
are presented in Appendix D. They
are taken from the analysis
of high energy photoproduction data made in Ref. \cite{10}.
\newline
(b) Resonance contributions are parametrized in the standard
Breit-Wigner form presented in Appednix C.
With increasing energy these  contributions
tend to 0, and the amplitudes of the modified  UIM 
are equal approximately to the Regge-pole
amplitudes beginning with  $W=2.5~GeV$.

Within the   unitary 
isobar model modified in this way, good description of
all multipole amplitudes with $l\leq 3$ is obtained
up to   $W=2~GeV$, taking $s_0=1.2~GeV^2$.
The results obtained using $\pi N$ amplitudes 
in the unitarization procedure (\ref {15}) from the 
GWU(VPI) analysis (SAID program) 
are presented in Figs. 1b-d and 2 (bold solid curves). 
For comparison in Figs. 1b and  2 the results obtained
with the same resonance contributions and with
the background of    the   UIM 
of Ref. \cite{1} 
are presented (thin solid curves). It is seen that for the multipoles
 ${}_pM_{1-}^{\frac{1}{2}}$,
${}_pM_{2-}^{\frac{1}{2}}$,
${}_pE_{2-}^{\frac{1}{2}}$,
${}_nE_{2-}^{\frac{1}{2}}$,
$E_{2-}^{\frac{3}{2}}$,
${}_pE_{3-}^{\frac{1}{2}}$,
${}_pM_{3-}^{\frac{1}{2}}$,
$M_{3+}^{\frac{3}{2}}$ 
they strongly contradict  the GWU(VPI) results.   

From Figs. 1b-d and 2a-i it is seen that 
for the multipole amplitudes with $l=0,1$
the deviation
between the results obtained with the modified
and nonmodified backgrounds becomes to be seen
at $W>1.3~GeV$, i.e. above first resonance region. 
For the multipoles with $l=2,3$,
except $E_{2+}^{3/2}$ (Fig. 2o), this deviation 
is noticeable already in the first resonance region.
However, for large multipoles
 ${}_pE_{2-}^{1/2},~{}_nE_{2-}^{1/2},~E_{2-}^{3/2}$ (Figs. 2l-n)
the deviation does not exceed the errors of the
GWU(VPI) data. For 
the multipoles
${}_pM_{2-}^{1/2},~{}_pE_{3-}^{1/2},
~{}_pM_{3-}^{1/2},~M_{3+}^{3/2}$ 
(Figs. 2k,p-r) there is no results
of partial-wave analyses below $W=1.3~GeV$,
however, these multipoles are very small in this energy region.
Moreover, with increasing energy 
the multipoles
${}_pM_{2-}^{1/2},~{}_pE_{3-}^{1/2},
~{}_pM_{3-}^{1/2},~M_{3+}^{3/2}$ 
obtained with the modified background are in better
agreement with the GWU(VPI) results.

Let us note that for the $P_{33}(1232)$ resonance 
we have  modified the Breit-Wigner parametrization
given in Eq. (C.1), 
taking $\Gamma_{total}$
in the form  $\Gamma_{total}=\Gamma_{\pi}\frac{M^2}{s}$.
This gives good description of the ratio of the imaginary
and real parts of the amplitudes $M_{1+}^{3/2}$, 
$E_{1+}^{3/2}$, $S_{1+}^{3/2}$
in accordance with the Watson theorem.
Let us note that background contributions
into these multipoles, unitarized via Eq. (\ref{15}),
satisfy the Watson theorem.
In the case of the amplitude $M_{1+}^{3/2}$ 
which is known with great
accuracy, the following modifications are also made:
\newline
(a) At $W<1.3~GeV$ the right part of the Eg. (C.1) 
is multiplied by the factor $(W/M)^6$. 
\newline
(b) At $W>1.3~GeV$ the imaginary and real parts of the resonance
contribution 
are multiplied by the factors $I_{Im}$ and  $I_{Re}$:  
\begin{equation}
I_{Im}=\frac{(W/1.3)^{2.5}(1.3/M)^6}{1+2.4(s-1.69)^{2.5}},~~ 
I_{Re}=\frac{(W/1.3)^{3.5}(1.3/M)^6}{1+1.4(s-1.69)^{2.5}}, 
\label{17}\end{equation}
where $W$ and $M$ are in the $GeV$ units.
These are slight modifications.
The obtained resonance contribution into $M_{1+}^{3/2}$
is presented in Figs. 1c,d by dashed-dotted curves.
It is seen that sum of this contribution and modified background
contribution (dashed curves) describe well the GWU(VPI) data.
With bakground of Ref. \cite{1} (dotted curves), 
resonance contribution should be strongly deformed
in order to describe  the GWU(VPI) data.

With the above described modifications,
we have obtained good description of the 
$P_{33}(1232)$ resonance contribution at $Q^2\neq 0$ too.
To demonstrate this in Fig. 4 the imaginary
part of $M_{1+}^{3/2}$ is presented at $Q^2=0.9,~1.8,~2.8,~4~GeV^2$
(solid curves).
In  this Figure  $Im~M_{1+}^{3/2}$ obtained within DR
via solution of the integral equation (\ref {8})
is also presented (dashed curves).
These results are obtained from the analysis
of JLab data \cite{18,19}.
It is seen that the amplitudes obtained
within two approaches are in good agreement with each other.
In order to compare the shape of the amplitude $M_{1+}^{3/2}$
at $Q^2\neq 0$ with its shape 
at $Q^2=0$, we have presented in   Fig. 4
the GWU(VPI) data with the normalizations corresponding
to  $Q^2=0.9,~1.8,~2.8,~4~GeV^2$. 
It is seen that the shape of $M_{1+}^{3/2}$
practically does not changed with increasing $Q^2$.

\section{Conclusion}

We have obtained good description of real parts
of all multipole amplitudes with $l\leq 3$
up to $W=1.8~GeV$ using fixed-t dispersion relations.
In Sec. 2.1 it is shown, that dispersion relations 
for multipole amplitudes $M_{1+}^{3/2}$,
$E_{1+}^{3/2}$ which correspond to the $P_{33}(1232)$ resonance,
can be transformed into singular integral equations.
The real and imaginary parts of these multipoles
were obtained via solution of these equations
and are in good agreement with the GWU(VPI) results.

We have modified the unitary isobar model of
Ref. \cite{1} via incorporation of Regge poles with increasing
energy and using unified Breit-Wigner parametrization
of resonance contributions in the form proposed
in Ref. \cite{7}. Within this approach
we have obtained good description of all photoproduction
multipoles with  $l\leq 3$
up to $W=2~GeV$.

Both approaches can be continued to  $Q^2\neq 0$,
and all formulas  in this paper are presented
in the form which allows to make this continuation.
Therefore,
both approaches can be used for analysis
of data on pion electroproduction on nucleons
and for extraction from these data information
on $Q^2$-evolution of $\gamma^* N \rightarrow N^*$
form factors. It is known, that 
this problem is very actual  at present
due to the experimental investigation
of this reaction on high duty-factor electron
accelerator at Jefferson Lab.
\begin{center}
{\large {\bf {Acknowledgments}}}
\end{center}

I am thankful to V.D.Burkert, who has drawn my attention
to the unitary isobar model. I am greatful to S.G.Stepanyan
for his cooperation in writing codes for both approaches.
I would like to thank V.I.Mokeev
for very usefull discussions, to
R.L.Crawford for 
discussions of the problems which can arise
in dispersion relations with increasing energy
and to I.I.Balitsky and N.Ya.Ivanov for discussions of the dynamics of 
reggezation
of amplitudes in the quark-gluon picture. 
I express my gratitude for the hospitality at Jefferson Lab
where this work was done.
 \vspace{1cm}

{\Large \bf Appendix A. Relations between invariant 
and multipole amplitudes}
\vspace{0.3cm}
\renewcommand\theequation{A.\arabic{equation}}
\setcounter{equation} 0

In order to connect invariant and multipole amplitudes
it is convenient to introduce
the intermediate amplitudes $f_i(s,cos \theta,Q^2)$: 
\begin{eqnarray}
f_1=&&\left[(W-m_N)B_1-B_5\right]\frac
{\left[(E_1+m_N)(E_2+m_N)\right]^{1/2}}{8\pi W},\nonumber \\
f_2=&&\left[-(W+m_N)B_1-B_5\right]\frac
{\left[(E_1-m_N)(E_2-m_N)\right]^{1/2}}{8\pi W}, \\
f_3=&&\left[2B_3-B_2+(W+m_N)\left(\frac{B_6}{2}-B_8 \right)\right]
\frac{\left[(E_1-m_N)(E_2-m_N)\right]^{1/2} 
(E_2+m)}{8\pi W},\nonumber \\
f_4=&&\left[-(2B_3-B_2)+(W-m_N)\left(\frac{B_6}{2}-B_8\right)\right]
\frac{\left[(E_1+m_N)(E_2+m_N)\right]
^{1/2}(E_2-m)}{8\pi W},\nonumber\\
f_5=&&\left\{\left[Q^2B_1+(W-m_N)B_5
+2W(E_1-m_N)\left(B_2-\frac{W+m_N}{2}B_6
\right)\right](E_1+m_N)\right.\nonumber \\
&&\left.-X\left[(2B_3-B_2)+(W+m_N)
\left(\frac{B_6}{2}- B_8 \right)\right]\right\}
\frac{(E_1-m_N)(E_2+m_N)}{8\pi WQ^2},\nonumber \\
f_6=&&\left\{-\left[Q^2B_1-(W+m_N)B_5
+2W(E_1+m_N)\left(B_2+\frac{W-m_N}{2}B_6
\right)\right](E_1-m_N)\right.\nonumber \\
&&\left.+X\left[(2B_3-B_2)-(W-m_N)
\left(\frac{B_6}{2}- B_8 \right) \right]\right\}
\frac{(E_1+m_N)(E_2-m_N)}{8\pi WQ^2},\nonumber
\end{eqnarray}
where
\begin{equation}
X=\frac{\tilde{k}_0}{2}(t-m_{\pi} ^2+Q^2)-Q^2\tilde{q}_0,
\end{equation}
$\theta$ is the polar angle of the pion in the c.m.s.,
$\tilde{k}_0,\tilde{q}_0,E_1,E_2$ are the energies
of virtual photon, pion, initial and final nucleons
in this system.

The expansion of the intermediate amplitudes over multipole
amplitudes $M_{l\pm}(s,Q^2)$, $E_{l\pm}(s,Q^2)$, 
$S_{l\pm}(s,Q^2)$
has the form:
\begin{eqnarray}
&&f_1=\sum\left\{(lM_{l+}+E_{l+})P'_{l+1}(cos\theta)+
\left[(l+1)M_{l-}+E_{l-} \right]
P'_{l-1}(cos\theta)\right\},\nonumber \\
&&f_2=\sum\left[(l+1)M_{l+}+lM_{l-} \right] P'_l(cos\theta),\nonumber \\
&&f_3=\sum\left[(E_{l+}-M_{l+})P''_{l+1}(cos\theta)+
(E_{l-}+M_{l-})P''_{l-1}(cos\theta)\right],\\
&&f_4=\sum(M_{l+}-E_{l+}-M_{l-}-E_{l-})P''_l(cos\theta),\nonumber \\
&&f_5=\sum\left[(l+1)S_{l+}P'_{l+1}(cos\theta)-
lS_{l-}P'_{l-1}(cos\theta)\right],\nonumber \\
&&f_6=\sum\left[lS_{l-}-(l+1)S_{l+}\right] P'_l(cos\theta).\nonumber
\end{eqnarray}
The formulas which relate the amplitudes 

$f_i(s,cos \theta,Q^2)$ to the helicity
amplitudes and cross section can be found in Ref. (\cite{14}).
\vspace{1cm}

{\Large \bf Appendix B. Born contribution}
\vspace{0.3cm}
\renewcommand\theequation{B.\arabic{equation}}
\setcounter{equation} 0

The residues in the nucleon poles of the invariant
amplitudes in Eqs. (\ref{3},\ref{4}) are equal to:
\begin{eqnarray}
&&R_1^{(v,s)}(Q^2)=\frac{ge}{2}\left(F_1^{(v,s)}(Q^2)
+2mF_2^{(v,s)}(Q^2)\right),\nonumber \\
&&R_2^{(v,s)}(Q^2)=-\frac{ge}{2}F_1^{(v,s)}(Q^2),\nonumber \\
&&R_3^{(v,s)}(Q^2)=-\frac{ge}{4}F_1^{(v,s)}(Q^2), \\
&&R_5'^{(v,s)}(Q^2)=\frac{ge}{4}
(m_{\pi^2} -Q^2-t)F_2^{(v,s)}(Q^2),\nonumber \\
&&R_6^{(v,s)}(Q^2)=ge F_2^{(v,s)}(Q^2),\nonumber \\
&&R_8^{(v,s)}(Q^2)=\frac{ge}{2}F_2^{(v,s)}(Q^2),\nonumber
\end{eqnarray}
where $g^2/4\pi=14.2,~e^2/4\pi=1/137$, and
$F_1^{(v,s)}(Q^2),~F_2^{(v,s)}(Q^2)$ are the nucleon Pauli
form factors.
Following Ref.\cite {1}, in the UIM  
the Lagrangian for the  $\pi N N$ vertex 
is taken in the form of mixed pseudovector (PV) 
and pseudoscalar (PS) couplings:
\begin{equation}
L_{\pi N N}=\frac{\Lambda^2}{\Lambda^2+q^2}L_{\pi N N}^{PV}+
\frac{q^2}{\Lambda^2+q^2}L_{\pi N N}^{PS},
\end{equation}
where we take $\Lambda^2=0.12~GeV^2$.
This leads to the following additional contributions 
in the amplitudes $B_1^{(+,0)}(s,t,Q^2)$:
\begin{equation}
B_1^{(+,0)}(s,t,Q^2)=B_1^{(+,0)}(s,t,Q^2)+AF_2^{(v,s)}(Q^2),
\end{equation}
where
\begin{equation}
A=\frac{ge}{2 m_N}\frac{\Lambda^2}{\Lambda^2+q^2}.
\end{equation}
 The nucleon Pauli form factors $F_1^{(v,s)}(Q^2),~F_2^{(v,s)}(Q^2)$
 in the above equations we have defined 
according to the description of the existing data in Refs.\cite{20,21,22} 
in the following way:
\begin{eqnarray}
&&F_1^{(v,s)}(Q^2)=F_{1p}(Q^2)-F_{1n}(Q^2),
~F_2^{(v,s)}(Q^2)=F_{2p}(Q^2)-F_{2n}(Q^2),\nonumber\\
&&F_{1p}(Q^2)=G_p^m(Q^2)/(1+2m_Nz),~~F_{2p}(Q^2)=zF_{1p}(Q^2),\nonumber\\
&& z=\frac{1.793}{2m_N}(1+\frac{1.2Q^2}{1+1.1Q}+0.015Q^2+0.001Q^8),\\
&&G_p^m(Q^2)=2.793/(1+0.35Q+2.44Q^2+0.5Q^3+1.04Q^4+0.34Q^5),\nonumber\\
&&F_{1n}(Q^2)=\frac{G_n^e(Q^2)+\tau G_n^m(Q^2)}{1+\tau},~~
F_{2n}(Q^2)=\frac{G_n^m(Q^2)-G_n^e(Q^2)}{2m_N(1+\tau)},\nonumber\\
&&\tau=Q^2/4m_N^2,~~
G_n^e=\frac{0.5Q^2}{1+25Q^4},~~   
G_n^m(Q^2)=-1.913F_d(Q^2),\nonumber\\
&&F_d(Q^2)=1/(1+0.71/Q^2).\nonumber
\end{eqnarray}
Here it is supposed that $Q^2$ is taken in the $GeV^2$ units.
The pion form factor we take as in Ref. \cite {1} in the form
\begin{equation}
F_\pi(Q^2)=F_1^{(v)}(Q^2).
\end{equation}
\vspace{1cm}

{\Large \bf Appendix C. Breit-Wigner parametrization 
for resonance contributions}
\vspace{0.3cm}
\renewcommand\theequation{C.\arabic{equation}}
\setcounter{equation} 0

We use
the Breit-Wigner parametrization 
for the resonance contributions into multipole amplitudes
in the form \cite{7,9}:
\begin{equation}
Res_{B-W}(W,Q^2)=c(\frac{k}{k_r})^n
\left(\frac{q_r }{q}\frac{k_r}{k}\frac{\Gamma_\pi \tilde{\Gamma}_\gamma }
{\eta_\pi \Gamma }\right)^{1/2}
\frac{M\Gamma}{M^2-W^2-iM\Gamma_{total}},
\end{equation}
where $n=0$ for $M_{l\pm},~E_{l\pm}$,  $n=1$ for $S_{l\pm}$ and
\begin{eqnarray}
  &&\Gamma_{total}=\Gamma_{\pi}+\Gamma_{inel},\\
 &&\Gamma_{\pi}=\eta_{\pi}\Gamma\left(\frac{q}{q_r}\right)^{2l+1}
 \left(\frac{X^2+q_r^2}{X^2+q^2}\right)^l,\\
 &&\tilde{\Gamma}_{\gamma}=\left(\frac{k}{k_r}\right)^{2l^{\prime}+1}
 \left(\frac{X^2+k_r^2}{X^2+k^2}\right)^{l^{\prime}},\\
 &&\Gamma_{inel}=
(1-\eta_{\pi})\Gamma\left(\frac{q_{2\pi}}{q_{2\pi,r}}\right)^{2l+4}
 \left(\frac{X^2+q_{2\pi,r}^2}{X^2+q_{2\pi}^2}\right)^{l+2}. 
\end{eqnarray}
For $M_{l\pm},E_{l+},S_{l+}$ $l'=l$,
for $E_{l-},S_{l-}~l'=l-2$ if $l\geq 2$,
for $S_{1-}~l'=1$; 
$M$ and $\Gamma$ are masses and widths of the resonances,
$\eta_{\pi}$ are the branching ratios into the $\pi N$ channel,
$q_{2\pi}$ is the 3-momentum of the $2\pi$ system
in the decay $Resonance\rightarrow 2\pi+N$ in the c.m.s., 
$q_{2\pi,r}$ is the magnitude of this momentum at $W=M$,
$X$ are phenomenological parameters.

For the resonance $S_{11}(1535)$ which has 
large branching ratio into the
$\eta N$ channel, $\Gamma_{total}$ is taken in the form
\begin{equation}
\Gamma_{total}=0.6\Gamma_\pi+0.1\Gamma_{inel} 
+0.3\Gamma\frac{q_\eta}{q_\eta^r}.
\end{equation}
Below the thresholds 
of $2\pi+N$ and  $\eta+N$ productions we take, 
respectively, $q_{2\pi}=0$ and $q_\eta=0$.
\vspace{1cm}

{\Large \bf Appendix D. Invariant amplitudes in the \\ Regge regime}
\vspace{0.3cm}
\renewcommand\theequation{D.\arabic{equation}}
\setcounter{equation} 0

In Ref. \cite{10} the introduction of Regge-trajectories
is made in gauge-invariant form
for invariant amplitudes in the following way.
Instead of t-channel $\pi$ exchange, which is
non-gauge-invariant
and contributes into $B_3^{(-)}(s,t,Q^2)$, 
the following combination is used
\begin{equation}
\bar{u}(p_2)\gamma _5 \left\{2P ^\mu
B_2^{(-)}+2q^\mu B_3^{(-)}\right\} u(p_1),
\end{equation}
where in addition to the $\pi$ contribution, nucleon pole contribution
generated by the form factor $F_1^{(v)}(Q^2)$ is taken into account.
The nucleon and pion contributions into
$B_2^{(-)}$ and $B_3^{(-)}$ are written
in the form:
\begin{eqnarray}
&&B_2^{(-)}(s,t,q^2)=-\frac{ge}{2}F_1^{(v)}(Q^2)
\left(\frac{t-m_\pi^2}{s-m^2}-\frac{t-m_\pi^2}{u-m^2}\right)
\frac{1}{t-m_{\pi}^2}, \\
&&B_3^{(-)}(s,t,q^2)=\left[-\frac{ge}{4}F_1^{(v)}(Q^2)
\left(\frac{t-m_\pi^2}{s-m^2}+\frac{t-m_\pi^2}{u-m^2}\right)
-egF_\pi (Q^2)\right]
\frac{1}{t-m_{\pi}^2},
\end{eqnarray}
and are reggezied by the replacement:
\begin{equation}
\frac{1}{t-m_{\pi}^2}~\Rightarrow~P^{\pi}_{Regge}~in~B_{2,3}^{(-)}
~+~P^{b_1}_{Regge}~in~B_{2,3}^{(0)}.
\end{equation}
Here $P_{Regge}$ are Regge propogators: 
\begin{equation}
P^{\pi, b_1}_{Regge}=\left(\frac{s}{s_0}\right)^{\alpha_\pi (t)}
\frac{\pi \alpha'_\pi}{sin(\pi \alpha_\pi (t))}
\frac{1}{\Gamma(1+\alpha_\pi (t))}
\frac{\tau+e^{-i\pi \alpha_\pi (t))}}{2},
\end{equation}
and it is supposed that $\pi$ and $b_1$ trajectories
are degenerate; $\alpha_\pi (t)=\alpha'_\pi(t-m_\pi^2),
~ \alpha'_\pi=0.7~GeV^2$.

The contributions of the $\rho$ and $\omega$ exchanges
in the $t$-channel are gauge invariant and are reggezied
simply by the following replacements in Eqs. (E.1):
\begin{eqnarray}
&&\frac{1}{t-m_{\rho}^2}~\Rightarrow~
P^{\rho}_{Regge}~in~B_i^{(0)}
~+~P^{a_2}_{Regge}~in~B_i^{(-)},~~i=1,2,3,6, \\
&&\frac{1}{t-m_{\omega}^2}~\Rightarrow~
P^{\omega}_{Regge}~in~B_i^{(+)},~~i=1,2,3,6,
\end{eqnarray}
where it is supposed that $\rho$ and $a_2$ trajectories
are degenerate and
\begin{eqnarray}
&&P^{\rho ,a_2}_{Regge}=
\left(\frac{s}{s_0}\right)^{\alpha_\rho (t)-1}
\frac{\pi \alpha'_\rho}{sin(\pi \alpha_\rho (t))}
\frac{1}{\Gamma(\alpha_\rho (t))}
\frac{\tau+e^{-i\pi \alpha_\rho (t))}}{2},\nonumber \\ 
&&P^{\omega }_{Regge}=
\left(\frac{s}{s_0}\right)^{\alpha_\omega (t)-1}
\frac{\pi \alpha'_\omega}{sin(\pi \alpha_\omega (t))}
\frac{1}{\Gamma(\alpha_\omega (t))}
\frac{\tau+e^{-i\pi \alpha_\omega (t))}}{2},
\end{eqnarray}
$\alpha_\rho (t)=0.55+\alpha'_\rho t,~ 
\alpha'_\rho=0.8~GeV^2,~
\alpha_\omega (t)=0.44+\alpha'_\omega t,~ 
\alpha'_\omega=0.9~GeV^2$.
In Eqs. (D.5,D.8) $\tau$ is the signature
of the trajectory:
\begin{equation}
\tau_{\pi}=\tau_{a_2}=1,~\tau_{\rho}=\tau_{\omega}=\tau_{b_1}=-1.
\end{equation}
\vspace{1cm}

{\Large \bf Appendix E. $\rho$ and $\omega$ contributions}
\vspace{0.3cm}
\renewcommand\theequation{E.\arabic{equation}}
\setcounter{equation} 0

The $\rho$ and $\omega$ exchanges in the $t$ - channel contribute 
to the following amplitudes:
\begin{eqnarray}
&&B_1^{(0)}=\frac{e\lambda _\rho}{m_\pi}
\left[2m_Ng_{\rho 1}+t\frac{g_{\rho 2}}
{2m_N}\right]\frac{1}{t-m_{\rho}^2},\nonumber\\
&&B_2^{(0)}=\frac{e\lambda _\rho}{m_\pi}
\frac{g_{\rho 2}}{4m_N}
(Q^2+m_{\pi}^2-t)\frac{1}{t-m_{\rho}^2},\nonumber\\
&&B_3^{(0)}=\frac{e\lambda _\rho}{m_\pi}
\frac{g_{\rho 2}}
{8m_N}(u-s)\frac{1}{t-m_{\rho}^2},~~
B_6^{(0)}=2\frac{e\lambda _\rho}
{m_\pi}g_{\rho 1}\frac{1}{t-m_{\rho}^2},\\
&&B_1^{(+)}=\frac{e\lambda _\omega}{m_\pi}
\left[2m_Ng_{\omega 1}+t\frac{g_{\omega 2}}
{2m_N}\right]\frac{1}{t-m_{\omega}^2},\nonumber \\
&&B_2^{(+)}=\frac{e\lambda _\omega}{m_\pi}
\frac{g_{\omega 2}}{4m_N}(Q^2
+m_{\pi}^2-t)\frac{1}{t-m_{\omega}^2},\nonumber\\
&&B_3^{(+)}=\frac{e\lambda _\omega}{m_\pi}
\frac{g_{\omega 2}}{8m_N}(u-s)\frac{1}{t-m_{\omega}^2},~~
B_6^{(+)}=2\frac{e\lambda _\omega}{m_\pi}
g_{\omega 1}\frac{1}{t-m_{\omega}^2}.\nonumber
\end{eqnarray}
These amplitudes are obtained using verteces 
$\gamma \rho \pi,~\gamma \omega \pi,
~\rho NN$ and $\omega NN$ defined in the form presented
in Ref. \cite{1}. 
In the UIM, the off-shell behaviour 
of $g_{Vi}$ 
is described  by:
$g_{Vi}=\tilde{g}_{Vi}\Lambda_{V}^2/(\Lambda_{V}^2-t)$.
The coupling constants are taken from  Ref. \cite{1}
and are equal to:
\begin{eqnarray}
&&\lambda _\omega=0.314,~\tilde{g}_{\omega1}=21,
~\tilde{g}_{\omega2}=-12,~\Lambda_\omega=1.2,  \nonumber\\
&&\lambda _\rho=0.103,~\tilde{g}_{\rho1}=2,
~\tilde{g}_{\rho2}=13,~\Lambda_\rho=1.5.  
\end{eqnarray}
In the Regge-pole analysis of high energy
photoproduction data in  Ref. \cite{10} the coupling 
constants are:                                        
\begin{equation}
g^{Regge}_{\omega1}=13.9,
~g^{Regge}_{\omega2}=0,~
g^{Regge}_{\rho1}=3.47,~
g^{Regge}_{\rho2}=13.
\end{equation}

\newpage
{\Large \bf {Figure Captions}}
\vspace{1cm}
\newline
{\large \bf{Fig. 1}}. (a) Multipole amplitude $M_{1+}^{3/2}$
within DR.
The results are obtained 
via solution of the integral equation  (\ref{8}):
solid and dashed curves correspond to the 
real and imaginary parts of 
the amplitude, dotted and dashed-dotted curves
correspond to the contributions of homogeneous
and particular solutions into  $ImM_{1+}^{3/2}$, 
thin dashed curve is the contribution
of other multipoles into (\ref{8}). 
\newline
(b) Multipole amplitude $E_{1+}^{3/2}$ within DR and UIM.
Dashed-dotted and dashed  curves are 
the real and imaginary parts of
the amplitude  obtained by DR
via solution of the integral equation  (\ref{8}), 
thin dashed curve is the contribution
of other multipoles into (\ref{8});
solid and dotted  curves are 
the real and imaginary parts of
the amplitude  obtained within  UIM
with modified background,
thin solid curve is obtained
with the background of Ref. \cite{1}.
\newline
(c)   $ImM_{1+}^{3/2}$ in  the UIM:
solid curve is the full result
obtained with the modified background, dashed-dotted 
and dashed curves correspond to the resonance and background 
contributions, respectively,
dotted curve is 
the background of the UIM of Ref. \cite{1}.
\newline
(d)   $ReM_{1+}^{3/2}$ in  the UIM:
 the legend is as for Fig. 1c.
\newline
In Figs. 1a-d, the GWU(VPI) results from the 
SAID program are presented by open circles
for the imaginary parts of the amplitudes and
by solid triangles for those real parts.
\newline
{\large \bf{Fig. 2}}.  
Multipole amplitudes within
DR and UIM.
Dashed-dotted and dashed  curves are
the real and imaginary parts of
amplitudes  obtained by DR.
Solid and dotted  curves are
the real and imaginary parts of 
amplitudes  obtained within  UIM
with modified background,
thin solid curves are obtained   
with the background  of Ref. \cite{1}.
Plotted are the multipole amplitudes
in millifermi units, in the parentheses the notations
of multipole amplitudes in SAID program are given.
The GWU(VPI) results from the
SAID program  are presented by open circles
for the real parts of the amplitudes and 
by solid triangles for those imaginary parts.
\newline
{\large \bf{Fig. 3}}. The contributions of the integral
over $M_{1+}^{\frac{3}{2}}$ (in millifermi units) into the real parts   
of the multipole amplitudes  $E_{0+}^{\frac{1}{2}}$ 
(dashed curve),   
$E_{0+}^{\frac{3}{2}}$  (dashed-dotted curve),
 $M_{1-}^{\frac{1}{2}}$  (solid curve),
 $M_{1-}^{\frac{3}{2}}$  (dotted curve)
within dispersion relations.
\newline
{\large \bf{Fig. 4}}. The imaginary part of  
$M_{1+}^{\frac{3}{2}}$
obtained within DR (dashed curves)
and UIM  (solid curves) at 
$Q^2=0,~0.9,~1.8,~2.8,~4~GeV^2$. Data at $Q^2=0$ are
from GWU(VPI) analysis, data at other $Q^2$ are the
same data with changed normalizations.
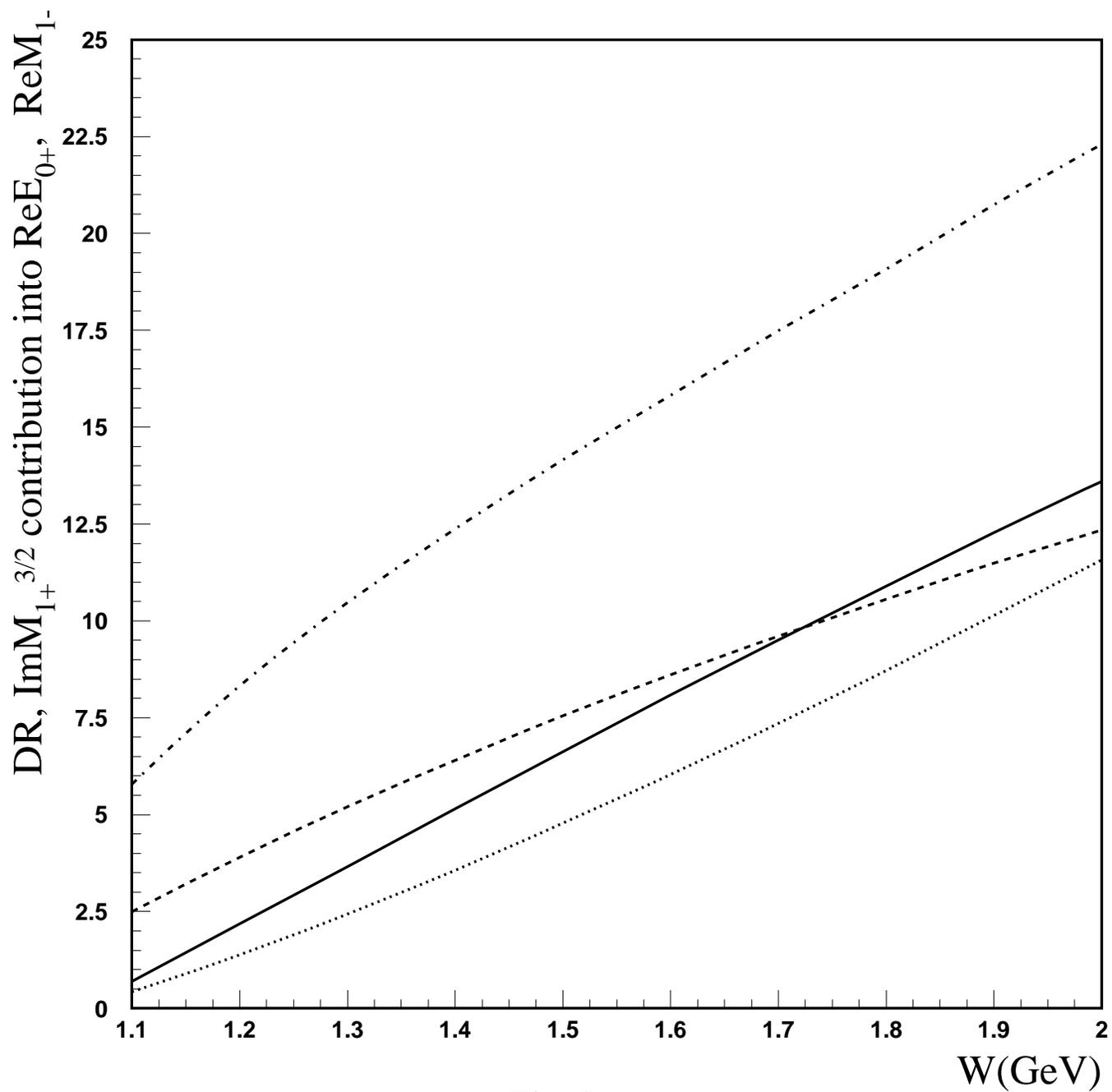
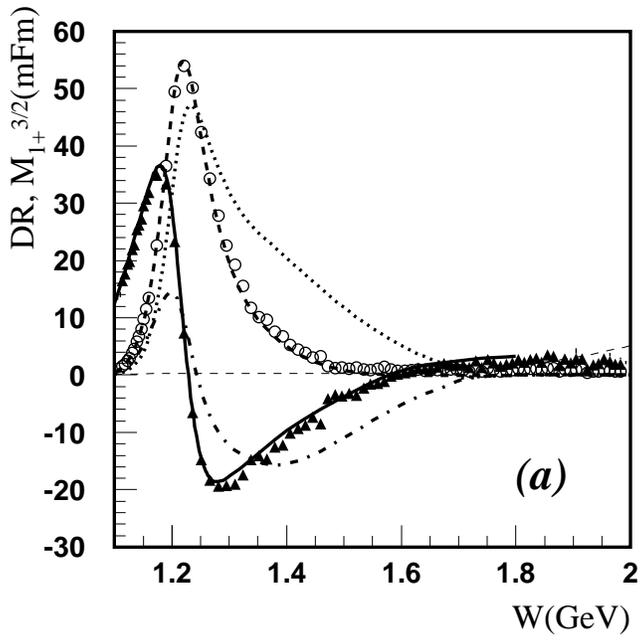
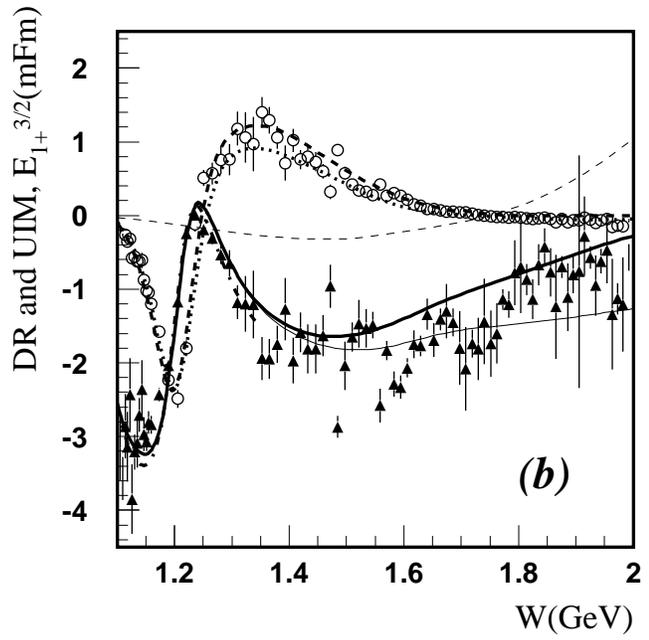
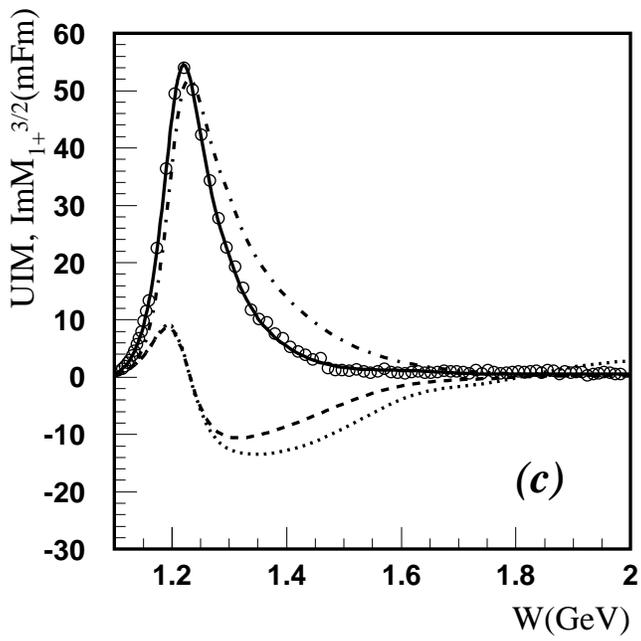
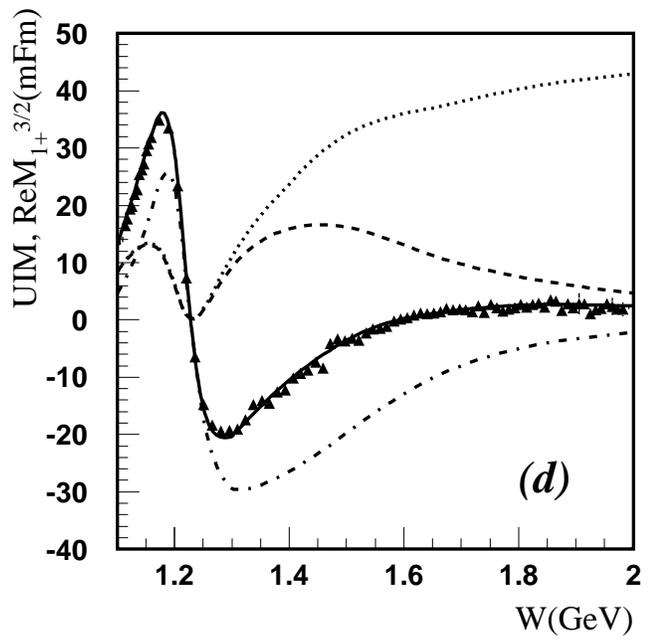
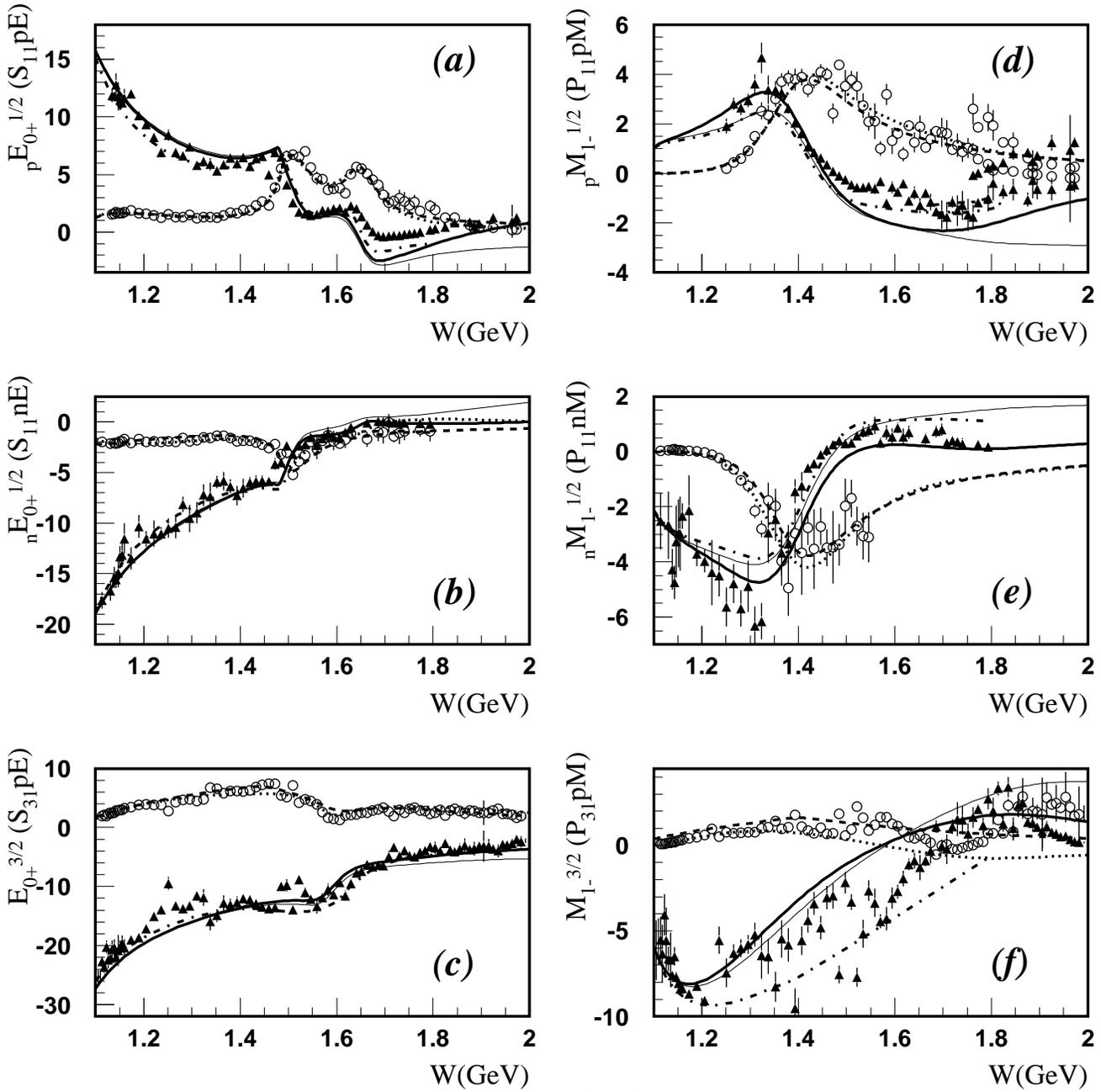
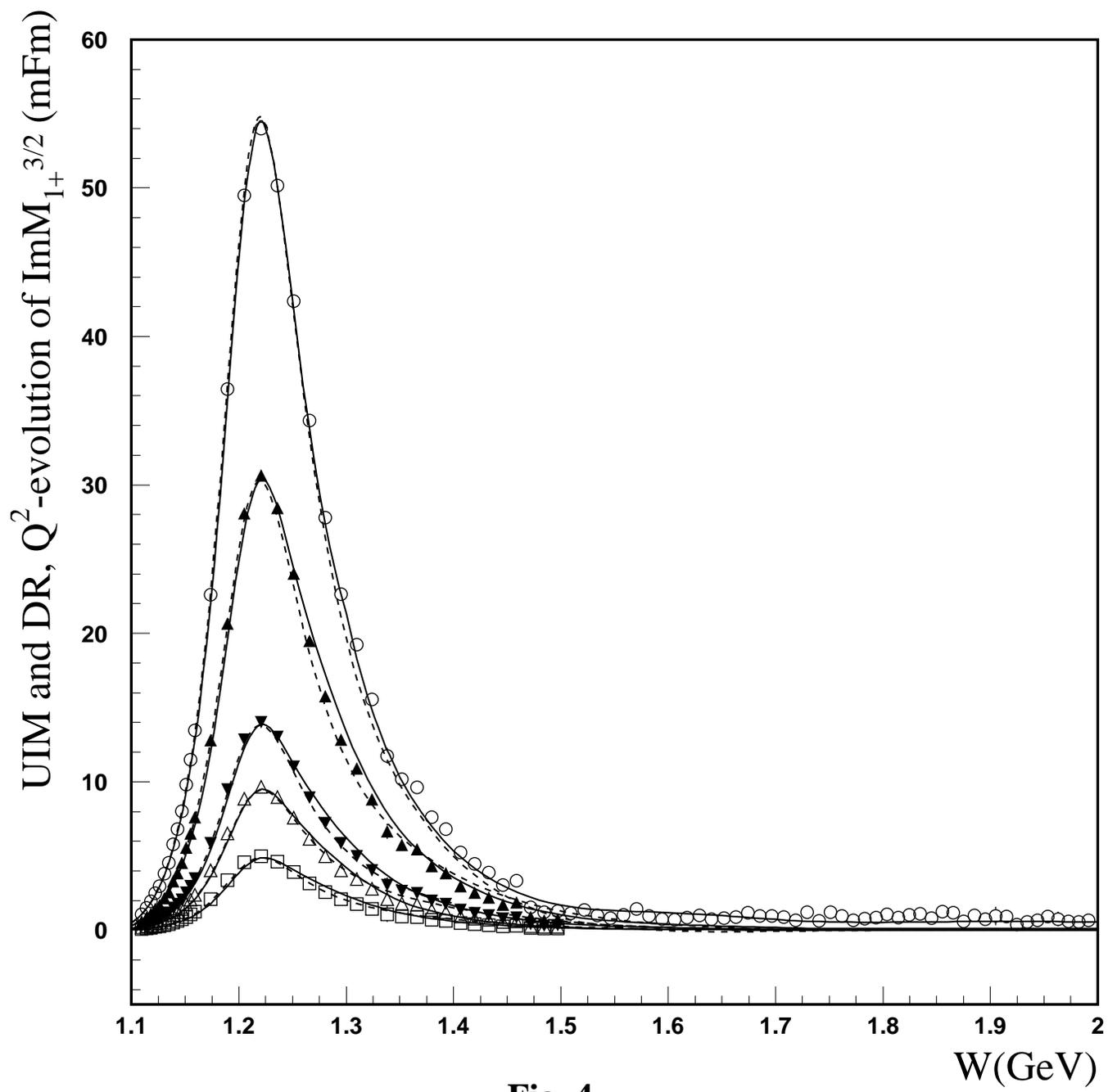
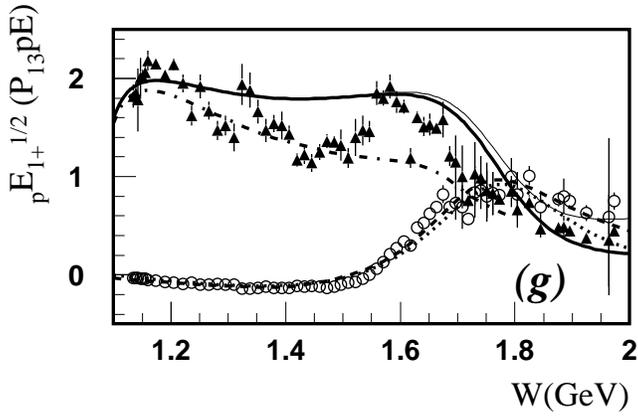
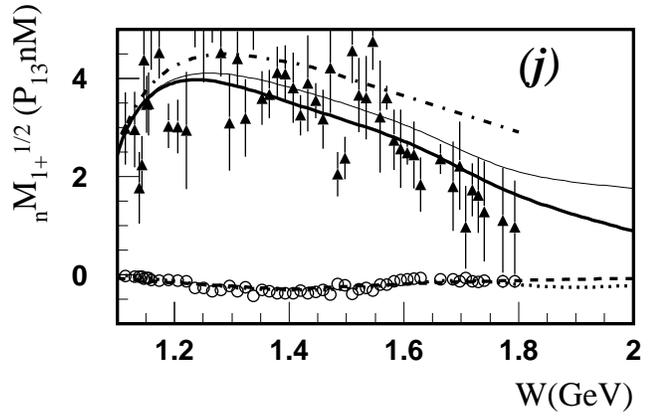
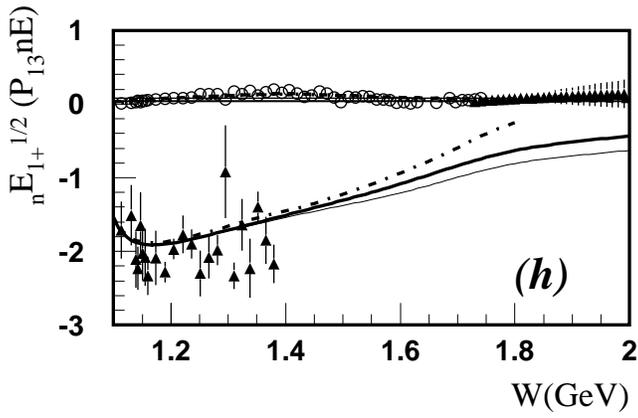
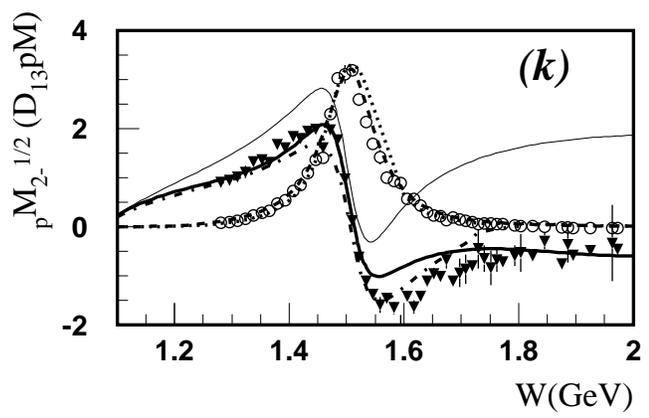
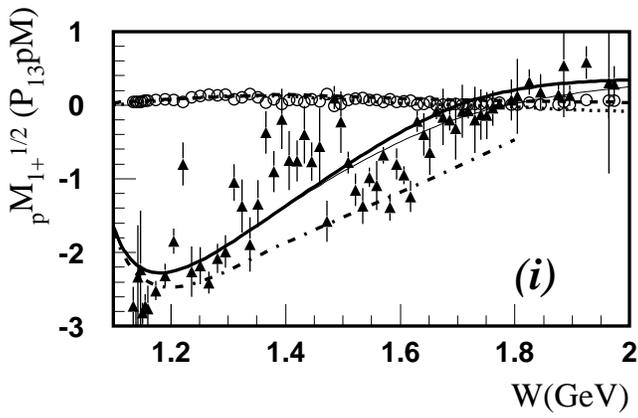
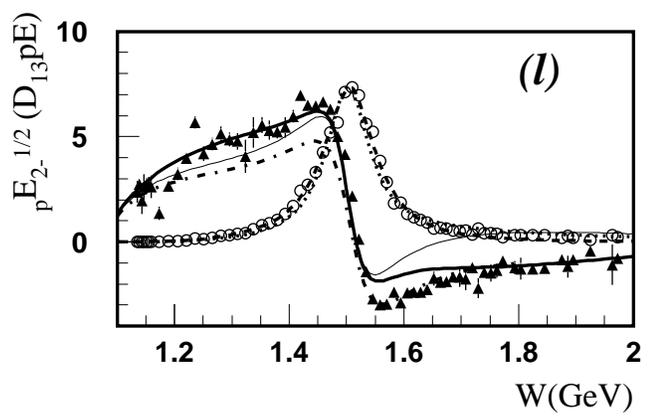
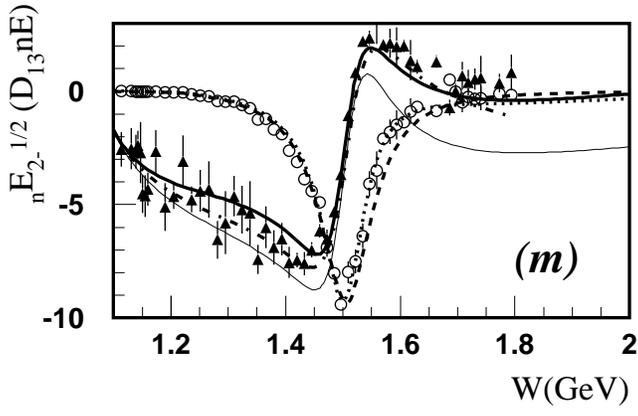
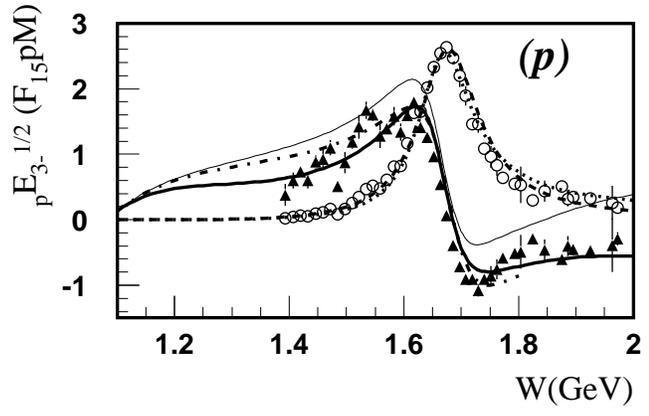
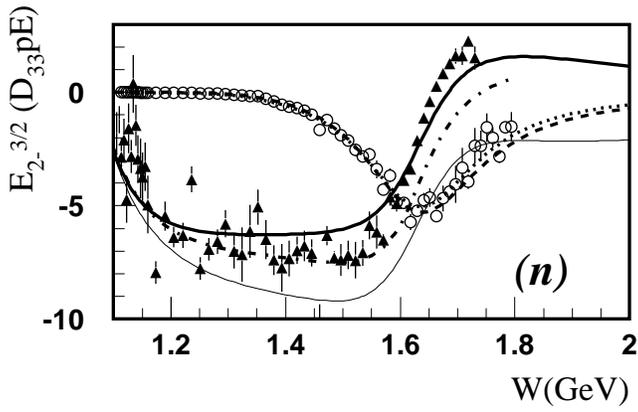
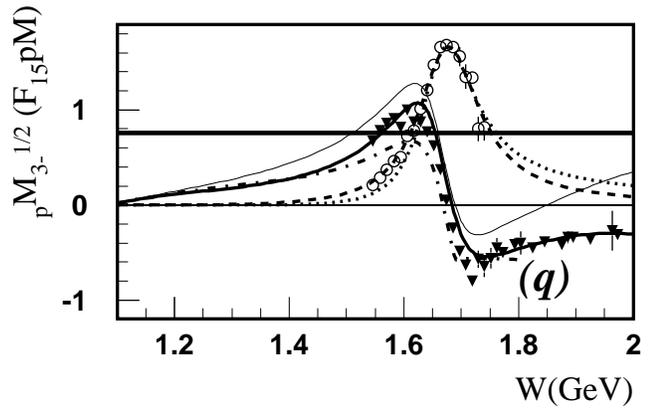
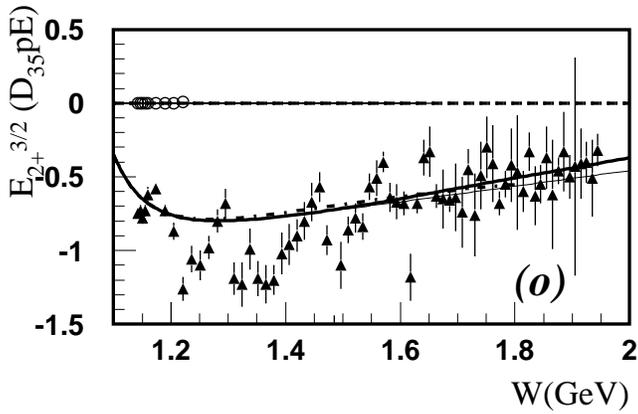
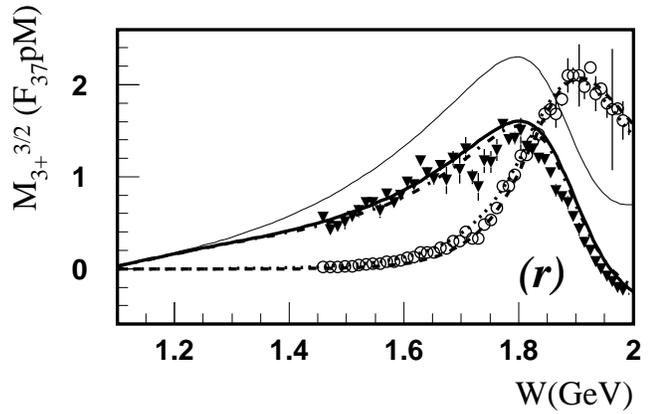
\end{document}